\documentstyle[11pt]{article}

\newcommand{\Title}[1]{{\large\bf\boldmath #1 \\[3mm] {\footnotesize by}
  \\[3mm]}}
\newcommand{\Author}[2]{{\large\spaceskip 2pt plus 1pt minus 1pt #1}\\[3mm]
  {\small #2}\\[6mm]}

\newcommand{\Abstract}[2]{{\footnotesize\begin{center}ABSTRACT\end{center}
  \vspace{1mm}\par#1\par
  \noindent {\bf Key words:~~}{\it #2}}}

\newcommand{\TabCap}[2]{\begin{center}\parbox[t]{#1}{\begin{center}
  \small {\spaceskip 2pt plus 1pt minus 1pt T a b l e}
  \refstepcounter{table}\thetable \\[2mm]
  \footnotesize #2 \end{center}}\end{center}}

\newcommand{\TableFont}{\footnotesize}
\newcommand{\MakeTable}[4]{\begin{table}[htb]\TabCap{#2}{#3}
  \begin{center} \TableFont \begin{tabular}{#1} #4
  \end{tabular}\end{center}\end{table}}
\newcommand{\MakeFrameTable}[5]{\begin{table}[htb]\TabCap{#2}{#3}
  \begin{center} \TableFont \begin{tabular}{#1}\hline\trule #4\\[1mm]
  \hline\trule #5\\[1mm]\hline
  \end{tabular}\end{center}\end{table}}
\newenvironment{references}%
{
\footnotesize \frenchspacing

\newcommand{\ApJ}{Astrophys.\ J.}

\newcommand{\AJ}{Astron.\ J.}

\newcommand{\Acta}{Acta Astr.}

\renewcommand{\and}{{\rm and }}
\section{{\rm REFERENCES}}
\sloppy \hyphenpenalty10000
\begin{list}{}{\leftmargin1cm\listparindent-1cm
\itemindent\listparindent\parsep0pt\itemsep0pt}}%
{\end{list}\vspace{2mm}}

\def\TYLDA{~}
\newlength{\DW}
\newcommand{\dw}{\hspace{\DW}}

\newcommand{\refitem}[5]{\item[]{#1} #2%
\def\REFARG{#3}\ifx\REFARG\TYLDA\else, {\it#3}\fi
\def\REFARG{#4}\ifx\REFARG\TYLDA\else, {\bf#4}\fi
\def\REFARG{#5}\ifx\REFARG\TYLDA\else, {#5}\fi.}

\newcommand{\etal}{{\it et al.\ }}
\newcommand{\trule}{\rule{0pt}{14pt}}
\newcommand{\Acknow}[1]{\par\vspace{5mm}{\bf Acknowledgements.} #1}
\newcommand{\eg}{{\it e.g.},\,}
\newcommand{\zdot}{\makebox[0pt][l]{.}}
\newcommand{\up}[1]{\ifmmode^{\rm #1}\else$^{\rm #1}$\fi}

\newcommand{\uph}{\up{h}}
\newcommand{\upm}{\up{m}}
\newcommand{\ups}{\up{s}}
\newcommand{\arcd}{\ifmmode^{\circ}\else$^{\circ}$\fi}
\newcommand{\arcm}{\ifmmode{'}\else$'$\fi}
\newcommand{\arcs}{\ifmmode{''}\else$''$\fi}

\begin{document}

\renewcommand{\thefootnote}{\fnsymbol{footnote}}

\begin{center}

\Title{The Optical Gravitational Lensing Experiment.\\
Follow-up Observations of the MACHO Microlensing Event \\
in the Galactic Bulge.%
{\rm \footnote{Based on observations obtained at the
Las Campanas Observatory of the Carnegie Institution of Washington.}}}

\Author{
{}~M.~~S~z~y~m~a~\'n~s~k~i$^1$,
A.~~U~d~a~l~s~k~i$^1$,
{}~J.~~K~a~\l~u~\.z~n~y$^1$,
{}~M.~~K~u~b~i~a~k$^1$,
{}~W.~~K~r~z~e~m~i~\'n~s~k~i$^2$
{}~~and~~~M.~~M~a~t~e~o$^3$}
{$^1$Warsaw University Observatory,
Al.~Ujazdowskie~4, 00--478~Warszawa, Poland\\
e-mail: (msz,udalski,jka,mk)@sirius.astrouw.edu.pl\\[4pt]
$^2$Carnegie Observatories, Las Campanas Observatory, Casilla~601,
La~Serena, Chile\\
e-mail: wojtek@roses.ctio.noao.edu\\[4pt]
$^3$Department of Astronomy, University of Michigan, 821~Dennison
Bldg., Ann Arbor, MI~48109--1090, USA\\
e-mail: mateo@astro.lsa.umich.edu
}

\end{center}

\Abstract{We present follow-up observations of a gravitational
microlensing candidate found by the MACHO collaboration in the Galactic
bulge. The photometric data cover a  period near the maximum of the
event and may be used to construct complete light curve. The position of the
lensed star in the color magnitude diagram suggests that it  is a
Galactic bulge star located on the subgiant branch.}{dark matter --
gravitational lensing -- Stars: low-mass, brown dwarfs}

\vspace{3mm}
The Optical Gravitational Lensing Experiment (OGLE) is a long term
observing project designed to detect large numbers of microlensing
events. The OGLE project began in April 1992. Observations are made with
the 1-m Swope telescope of the Las Campanas Observatory (LCO), operated
by the Carnegie Institution of Washington. The detailed description of
the project can be found in Udalski \etal (1992), Szyma\'nski and
Udalski (1993). After three seasons of observations  12 microlensing
events candidates have been found (Udalski \etal 1994a,b). Several
side-projects have also given important results regarding the structure
of the Galaxy (Paczy\'nski \etal 1994b,c, Stanek \etal 1994) and the
Sagittarius and Sculptor dwarf galaxies (Mateo \etal 1994, Ka{\l}u\.zny
\etal 1994).

In 1994 OGLE introduced the Early Warning System (EWS) allowing an early
detection of on-going microlensing (Paczy\'nski 1994a, Udalski \etal
1994b). Nearly ''on-line'' detection of the events is extremely important
making it possible to improve the photometric coverage of an
event, especially at maximum light, allowing accurate estimation of the
parameters of the lensing and resolving fine details of the phenomenon
(\eg binary lenses, planets around lensing stars, lensed  star disk
size). Early detection also allows other observers around the world to make
timely follow-up multicolor photometric and spectroscopic observations. The
EWS system was routinely used during the whole 1994 OGLE season
resulting with detection of two first real time microlensing events:
OGLE~\#11 and OGLE~\#12 (Udalski \etal 1994b).

The MACHO collaboration -- another group searching for microlensing
phenomena -- had also been working on an ''on-line'' detection system
(Bennett -- priv.\ comm.). On September 1st, 1994, at the end of the OGLE
1994 observing season, the MACHO group announced their first ''on-line''
detected microlensing event candidate (Alcock \etal 1994) -- an object
located in the direction of the Galactic bulge
($\alpha$=$17\uph59\upm49\zdot\ups6$, $\delta$=$-28\arcd10\arcm56\arcs$
(2000.0)). The star was reported to be constant at $V$=$18\zdot\upm8$,
$R$=$17\zdot\upm8$ ($V,R$ -- approximate standard colors) during 1993
and mid-1994, and had brightened by $0\zdot\upm85$ during the week
ending August~31, 1994  with no change of color index.

The star reported by the MACHO group is located in a region which was
not monitored by OGLE. We decided to add the new field into our schedule
and make follow-up observations of the MACHO event. Due to the schedule
of nights allocated to the OGLE project at LCO we were able to collect
the exposures on nights of September 1--6 and 14--15, 1994  which ended
our observing season. A total of 9 $I$-band and 2 $V$-band exposures
were taken in various seeing/weather conditions. Reductions were done
using a modified version of the DoPHOT photometry program (Schechter,
Mateo and Saha 1993) on $220\times220$ pixels subframes centered on the
object. Differential photometry of the candidate relative to two nearby
constant stars was derived. Instrumental magnitudes were tied to the
standard $VI$ system using calibration procedure as described in Udalski
\etal (1992). Transformation from the instrumental to the standard
system was recalculated using  observations of a few dozen standard
stars. It turned out to be very close to the 1992 season transformation
(Udalski \etal 1992). We estimate the absolute magnitude scale error to
be smaller than 0.03 mag.  Table~1 lists the positions and  $V$ and $V-I$
magnitudes of the comparison stars.  Photometric data for the MACHO event are
shown in Table~2.

\MakeFrameTable{|l|c|c|}{8cm}
{Position and magnitudes of the comparison stars}
{            &     comparison A           &    comparison B}
{RA$_{2000}$ & $17\uph59\upm49\zdot\ups5$ & $-28\arcd11\arcm15\arcs$ \\
DEC$_{2000}$ & $17\uph59\upm50\zdot\ups5$ & $-28\arcd10\arcm39\arcs$ \\
$V$          &    17.475                  &   17.574 \\
$V-I$        &  \dw2.557                  & \dw2.251
}
\MakeTable{|c|c|c|c|c|c|c|}{9cm}
{OGLE photometry of the MACHO Galactic bulge event}
{\hline\multicolumn{7}{|c|}{\trule \large $V$-band} \\[1mm]
\hline\trule
Frame no &   JD hel  &  Exp.  & $V$  &  Error &   $V-I$ & Error \\
         &  -2448000 & time [s] &    &        &         &       \\[1mm]
\hline\trule
mr8405 & 1600.51501 & 1000   & 17.568 & 0.043 & 1.920 &  0.094 \\
mr8570 & 1611.56265 & \dw780 & 18.695 & 0.046 & 2.007 &  0.058 \\[1mm]
\hline
\multicolumn{7}{|c|}{\trule \large $I$-band} \\[1mm]
\hline\trule
Frame no  &  JD hel  &   Exp. &   $I$ &  Error & \multicolumn{2}{|c|}{} \\
          & -2448000 & time [s] &     &        & \multicolumn{2}{|c|}{} \\[1mm]
\cline{1-5}\trule
mr8323 & 1598.50715 &900 & 15.610 &  0.029 & \multicolumn{2}{|c|}{} \\
mr8365 & 1599.50606 &720 & 15.548 &  0.024 & \multicolumn{2}{|c|}{} \\
mr8373 & 1599.60069 &720 & 15.610 &  0.028 & \multicolumn{2}{|c|}{} \\
mr8380 & 1599.70871 &900 & 15.557 &  0.104 & \multicolumn{2}{|c|}{} \\
mr8422 & 1600.69561 &900 & 15.646 &  0.084 & \multicolumn{2}{|c|}{} \\
mr8452 & 1601.50830 &780 & 15.808 &  0.084 & \multicolumn{2}{|c|}{} \\
mr8493 & 1602.50204 &780 & 15.846 &  0.022 & \multicolumn{2}{|c|}{} \\
mr8533 & 1610.52972 &901 & 16.649 &  0.028 & \multicolumn{2}{|c|}{} \\
mr8569 & 1611.54965 &910 & 16.690 &  0.036 & \multicolumn{2}{|c|}{} \\[1mm]
\hline
}

Fig.~1 presents the light curve of the event in the $I$-band and $V-I$
color.
Because we had neither magnitude nor color information for the object
at its ''constant'' phase and the star was still fading when the OGLE
season ended, we were unable to obtain a reliable fit of our data to the
theoretical microlensing light curve and thus to obtain the parameters
of the event. We adopted the MACHO preliminary theoretical fit (Bennett,
priv.\ comm. -- solid line in Fig.~1):  time of maximum brightness
$T_{max}\hbox{(JD hel.)}=2449599.8$, time scale (the Einstein radius /
transverse velocity) $t_0=9.85$ days,  magnification $A=3.55$, and
applied  magnitude offset to get the best fit to our data (the offset
yielded constant $I$ magnitude~ $I_0=16.92$). This procedure can be
justified  as the microlensing event should be achromatic.

Most of our data cover the period near the maximum of the event.
Although some of the photometric measurements have relatively large
errors due to either bad seeing or bad weather conditions, the data
generally confirm microlensing as the most likely interpretation of the
increase of brightness of the MACHO candidate. The light variations fit
the theoretical light curve very well, the  $V-I$ color seems to be
constant over the course of the phenomenon. Thus our data may be used to
complete the light curves obtained by other observers and to compute more
accurate lensing parameters. The event seems to be a typical
short-scale, moderate amplitude microlensing, similar to those observed
in the Galactic bulge before. Observations near the maximum do not show
any statistically significant deviation from the theoretical fit
suggesting a normal, point-mass microlensing.

Figure~2 shows the color-magnitude diagram (CMD) of the $15\times 15$
arcmins field centered on the MACHO candidate (this field has been
designated as GB5). The procedure of constructing the CMD was similar to
other Galactic bulge fields (Udalski \etal 1993). The position of
the lensed star is shown as a white star. The lensed star lies
on the subgiant branch of the CMD and is very likely  located in the
Galactic bulge.

The successful ''real-time'' discovery of the microlensing events both
by OGLE and MACHO alert systems proves that in spite of their very low
probability of occurrence these events can be detected in their early
phases and that world wide follow-up observing campaigns can be
successfully organized.

\bigskip

Photometry of OGLE microlensing events, as well as regularly updated
OGLE status report can be found over the Internet at host \\
{\it sirius.astrouw.edu.pl} (148.81.8.1), using the "anonymous ftp"
service (directory {\small\sf /ogle}, files {\small\sf README, ogle.status,
early.warning}). Information on the recent OGLE status is also
available via "World Wide Web"\\
WWW: {\it http://www.astrouw.edu.pl/.}

\Acknow{This project was supported with the NSF grants AST
9216494 to B.~Paczy\'nski, AST 9216830 to G.W. Preston and Polish KBN
grant PB~0450/P3/94/06 to A.~Udalski.}

\end{document}